\documentclass[a4paper]{jpconf}
\usepackage{graphicx}

\usepackage{subfig}
\label{key}
\usepackage[english]{babel}
\usepackage{amsmath, amsthm, amssymb}
\usepackage{bbm}
\usepackage{url}
\usepackage{ulem}
\usepackage{tikz-cd}
\usepackage{braket}
\usepackage{path}
\usepackage{nicefrac}
\usepackage{amssymb}
 \usetikzlibrary{shapes,arrows}
\usepackage{amsmath}
\usepackage{amsthm}
\usepackage{cancel}
\usepackage{indentfirst}
\usepackage{eucal}
\usepackage{appendix}
\usepackage[utf8]{inputenc}
\usepackage{geometry}
\usepackage{verbatim}
\usepackage{nameref}
\usepackage{alltt}
\usepackage{hyperref}
\usepackage{pgfplots}
\usepackage{dsfont}
\usepackage{appendix}
\usepackage[font=small, labelfont=bf]{caption} 

\newtheorem{Definition}{Definition}

\begin{document}
\title{Polynomial algebras from commutants: Classical and Quantum aspects of $\mathcal{A}_3$}

\author{Rutwig Campoamor-Stursberg$^1$ , Danilo Latini$^2$, Ian Marquette$^3$ and Yao-Zhong Zhang$^3$}

\address{$^1$ Instituto de Matem\'{a}tica Interdisciplinar and Dpto. Geometr\'{i}a y Topolog\'{i}a, UCM, E-28040 Madrid, Spain}

\address{$^2$ Via Appia Antica 2, 00045 Genzano di Roma RM, Italy}

\address{$^3$ School of Mathematics and Physics, The University of Queensland, Brisbane, QLD 4072, Australia}

\ead{rutwig@ucm.es, latini.phys@gmail.com, i.marquette@uq.edu.au, yzz@maths.uq.edu.au}

\begin{abstract}
We review some aspects of the Racah algebra $R(n)$, including the closure relations, pointing out their role in superintegrability, as well as in the description of the symmetry algebra for several models with coalgebra symmetry. The connection includes the generic model on the $(n-1)$ sphere. We discuss an algebraic scheme of constructing Hamiltonians, integrals of the motion and symmetry algebras. This scheme reduces to the Racah algebra $R(n)$ and the model on the $(n-1)$ sphere only for the case of specific differential operator realizations. We review the method, which allows us to obtain the commutant defined in the enveloping algebra of $\mathfrak{sl}(n)$ in the classical setting. The related $\mathcal{A}_3$ polynomial algebra is presented for the case $\mathfrak{sl}(3)$. An explicit construction of the quantization of the scheme for $\mathcal{A}_3$ by symmetrization of the polynomial and the replacement of the Berezin bracket by commutator and symmetrization of the polynomial relations is presented. We obtain the additional quantum terms. These explicit relations are of interest not only for superintegrability, but also for other applications in mathematical physics.
\end{abstract}

\section{Introduction}

The Racah algebra algebra finds its origin in the coupling theory of angular momenta \cite{leb65,bie84a}. It was only in the 80ties when it was recognized that the Racah algebra $R(3)$ can be studied in its own right \cite{gra88}. In the 90ties, several nonlinear algebraic structures started to be discovered in the context of non-equivalent two-dimensional superintegrable systems  \cite{bon94}. The Racah algebra and its generalizations were found to be key to understand the classification of superintegrable systems and their connection to special functions \cite{pos07, kre07,pos11,kal13,gen13}. However, the nature of the symmetry algebra of $N$-dimensional systems still remains a difficult problem, and only recently, with the discovery of $R(4)$ \cite{kal11} and higher rank Racah algebras $R(n)$ \cite{gab19,bie20}, further progress has been achieved. It has been pointed out that higher rank quadratic algebras also constitute a powerful tool for the algebraic derivation of spectra \cite{lia18,lat19,cor21,lat21}. In this context, it has been observed that all classical and quantum systems with coalgebra symmetry admit the Racah algebra as a subalgebra of their symmetry algebra \cite{lat21b}. These models include models on $N$-dimensional spaces of constant \cite{bal06} and non-constant curvature \cite{bal09}. The study of the higher rank Racah algebra $R(n)$ has been pursued in various directions \cite{cram22}, but their representation theory is still an ongoing problem, even for the rank-2 Racah algebra $R(4)$ \cite{cram22b}.

\medskip 
These works rely on explicit realizations of the generators of the Racah algebra. In contrast, an entirely algebraic construction was proposed in \cite{cor20,cam21,cam22}, based on commutants of Lie algebras, while in \cite{cam22b,cam23}, the approach was based on Poisson-Lie algebras. It was shown how the symmetric algebra in the dual space of the Poisson-Lie of $\mathfrak{sl}(n)$, endowed with the Berezin bracket and the centraliser of the Cartan subalgebra provide algebraic Hamiltonians, integrals and the symmetry algebra \cite{cam23}. This polynomial algebra, referred to as $A_n$ and of degree $(n-1)$, collapses onto $R(n)$ only for specific realizations, as well as the algebraic Hamiltonian collapses on the generic model on the $(n-1)$ sphere. 

\medskip 
The quantum analogue of the $A_n$ polynomial algebra was only described implicitly, and the aim of this paper is twofold: to review some aspects of the Racah algebra $R(n)$ and the construction of the algebra $A_n$ in the context of Poisson-Lie algebras, and to propose an explicit description of the quantum algebra $\mathcal{A}_3$. We plan to describe the polynomials in the enveloping algebra of $sl(3)$ explicitly, using the symmetrization map, in order to obtain the quantum correction for the polynomial algebra $\mathcal{A}_3$, which is of interest for applications in quantum physics and quantum superintegrable systems. This also allows to describe explicitly the classical/quantum (Poisson-Lie /Lie algebra) correspondence in regard of the construction of polynomial algebras.

\section{The Racah algebra $R(n)$}

The (classical/Poisson) higher rank Racah algebra $R(n)$ can be constructed via left and right Casimirs in the coalgebra with certain realizations of $\mathfrak{sl}(2)$ \cite{lat21b} and involves generators $P_{ij}$ (with $i,j=1,2,3,...n$)  and $F_{ijk}:=\frac{1}{2}\{P_{ij},P_{jk}\}$. We get a quadratic Poisson algebra, whose defining relations are those of the Racah algebra $R(n)$:
\begin{align}
&\{P_{ij},P_{jk}\}=2 F_{ijk}  \label{eq:PoissonRacah1}\\
&\{P_{jk}, F_{ijk}\}=P_{ik}P_{jk}-P_{jk}P_{ij}+2P_{ik}C_j-2P_{ij}C_k  \label{eq:PoissonRacah2} \\
&\{P_{kl}, F_{ijk}\}=P_{ik}P_{jl}-P_{il}P_{jk} \label{eq:PoissonRacah3} \\
&\{F_{ijk}, F_{jkl}\}=F_{jkl}P_{ij}-F_{ikl}(P_{jk}+2C_j)-F_{ijk}P_{jl} \label{eq:PoissonRacah4} \\
&\{F_{ijk}, F_{klm}\}=F_{ilm}P_{jk}-P_{ik}F_{jlm}
\label{eq:PoissonRacah5}
\end{align}
for $i,j,k,l,m,r \in \{1,\dots, n\}$. It is also understood that different indices take different values. For $n=3$, only the relations (\ref{eq:PoissonRacah1}) and (\ref{eq:PoissonRacah3}) appear. For $n \geq 4$, the relations (\ref{eq:PoissonRacah3}) and (\ref{eq:PoissonRacah4}) must be incorporated. Finally, for $n \geq 5$, an additional set of relations given by (\ref{eq:PoissonRacah5}) must be added. As $\{F_{ijk}, F_{lmn}\}=0$, no further relations are required, and the algebra is valid for any $n$.

As demonstrated in \cite{lat21b}, this is the subalgebra of all quasi-maximal superintegrable systems ($2n-2$ independent integrals with the Hamiltonian) with coalgebra. Among these we find 
\begin{equation}
H= p^2 + \sum_1^n \frac{\alpha_i}{x_i^2} + V(r),
\end{equation}
where $V(r)$ is any potential in the radial variable. This allows to view the $R(n)$ algebra as playing an analogous role for Hamiltonians with coalgebra as the Lie algebra $so(n)$ for systems with rotational invariance. The generic model on the sphere $\mathbb{S}^{n-1}$ has $R(n)$ as symmetry algebra. It is, provided constraints, the total Casimir and has $2n-1$ independent integrals. It is given by 
\begin{equation*}
H=\frac{1}{2}\sum_{1\leq i<j}\left(
s_{i}p_{j}-s_{j}p_{i}\right) ^{2}+\frac{1}{2}\sum_{k=1}^{n}\frac{\alpha
_{k}^{2}}{s_{k}^{2}},\quad \sum_{k=1}^{n}s_{k}p_{k}=0,\;\sum_{k=1}^{n}s_{k}^{2}=1.
\end{equation*}

The closure of the algebra (and quantum analog) was also obtained \cite{lat21b}:
\begin{equation}\label{eq:ho1}
{\footnotesize
\begin{split}
F_{ijk}^2& -C_i P_{jk}^2-C_j P_{ik}^2-C_k P_{ij}^2+P_{ij}P_{jk}P_{ik}+4 C_i C_j C_k=0,\\
2F_{ijk}&F_{jkl}-P_{il}P_{jk}^2+P_{ij}P_{jk}P_{kl}+P_{ik}P_{jk}P_{jl}-2C_j P_{ik}P_{kl}-2C_k P_{ij}P_{jl}+4C_jC_kP_{il}=0,\\
2F_{ijk}&F_{klm}-P_{il}P_{jk}P_{km}-P_{ik}P_{jm}P_{kl}+P_{im}P_{jk}P_{kl}+P_{ik}P_{jl}P_{km}-2C_kP_{im}P_{jl}+2C_kP_{il}P_{jm}=0,\\
2F_{ijk}&F_{lmr}-P_{il}P_{jr}P_{km}-P_{ir}P_{jm}P_{kl}-P_{kr}P_{im}P_{jl}+P_{im}P_{jr}P_{kl}+P_{ir}P_{jl}P_{km}+P_{il}P_{jm}P_{kr}=0.
\end{split} }
\end{equation}

\section{The algebraic setting in the Lie and Poisson-Lie context}

We briefly review the construction introduced in \cite{cam23} in regard of commutants, i.e., centralizers of a subalgebra of a Lie algebra in their enveloping algebra and their Poisson-Lie analogue related to the symmetric algebra. Explicitly, we define

\begin{Definition} 
The commutant $C_{\mathcal{U}(\mathfrak{s})}(\mathfrak{a})$ of a subalgebra $\mathfrak{a}\subset \mathfrak{s}$ is defined as the centralizer of $\mathfrak{a}$ in $\mathcal{U}(\mathfrak{s})$ : 
\begin{equation}
C_{\mathcal{U}(\mathfrak{s})}(\mathfrak{a})=\left\{ Q\in\mathcal{U}(\mathfrak{s})\; |\; [P,Q]=0,\quad \forall P\in\mathfrak{a}\right\}.\label{comm}
\end{equation}
\end{Definition}
The structure of the problem is quite complicated. More details on it can be found e.g. in \cite{cam22b,cam23}. In the case of $\mathfrak{sl}(n)$ and other semisimple or reductive algebras, the commutant $C_{\mathcal{U}(\mathfrak{s})}(\mathfrak{a})$ is in fact Noetherian and finitely generated. A basis of the commutant $C_{\mathcal{U}(\mathfrak{s})}(\mathfrak{a})$ is spanned, as vector space, by the elements  $P_1^{a_1}P_2^{a_2} \dots P_s^{a_s}, \quad a_i\in\mathbb{N}\cup {0}$, with additional constraints on the coefficients $a_i$. The canonical isomorphism $\Lambda$  allows us to look into the classical problem (Poisson-Lie algebra and Berezin bracket setting). In this context, the problem consists in determining the commutant of a subalgebra $\frak{a}$ of $\frak{s}$ in the enveloping algebra $\mathcal{U}\left( \frak{s}\right) $ in the Lie-Poisson context, i.e., for $\frak{a}^{\ast }\subset \frak{s}^{\ast }$. 
\begin{Definition}
The centralizer of $\mathfrak{a}$ in the symmetric algebra $S(\mathfrak{s})$ is defined as 
\begin{equation*}
C_{S(\frak{s})}\left( \frak{a}\right) =\left\{ Q\in S\left( 
\frak{s}\right) \;|\;\left\{ P,Q\right\} =0,\;P\in \frak{a}\right\} .
\end{equation*}
\end{Definition}
In this context, the problem can be recast into a system of partial differential equations, enabling us to use different methods, such as the methods of characteristics
\begin{equation}
\widehat{X_{i}}\left( Q\right) :=\left\{ x_{i},Q\right\} =C_{ij}^{k}x_{k}%
\frac{\partial Q}{\partial x_{j}}=0,\;1\leq i\leq m=\dim \frak{a},
\label{SIS}
\end{equation}
where $\left\{ x_{1},\ldots ,x_{m}\right\} $ are coordinates in a dual basis
of $\frak{a}^{\ast }$. This analytical reformulation allow us to make further progress in the explicit construction, and to get insight into the quantum case. One important ingredient in the construction of the quantum analogue is the use of the symmetrization map
\begin{equation}\label{syma}
\Lambda\left(x_{j_1}\dots x_{j_p}\right)=\frac{1}{p!} \sum_{\sigma\in  \Sigma_{p}} X_{j_{\sigma(1)}}\dots X_{j_{\sigma(p)}},
\end{equation}
As will be pointed out in the next section, this symmetrization map will be implemented in regard of the basis of polynomials which are linearly independent, but as well at the level of polynomial relations. We next provide details on the classical construction in regard of $C_{S(\mathfrak{sl}(3))}(\mathfrak{h} )$ and the Poisson polynomial algebra $\mathcal{A}_3$.

\section{$C_{S(\mathfrak{sl}(3))}(\mathfrak{h} )$ and the Poisson polynomial algebra $\mathcal{A}_3$}

We consider the special linear Lie algebra $\mathfrak{sl}(n)$ in its defining representation. Starting with a basis given by the generators $E_{ij}$ with $1\leq i,j\leq n$ subjected to the constraint $\sum_{i=1}^n E_{ii}=0$, we use the Poisson-Lie algebra, which is taken as (where $e_{ij}$ are the generators of the Poisson-Lie algebra, i.e. $E_{ij} \rightarrow e_{ij}$ and $[,] \rightarrow \{,\}$)

\begin{equation}\label{kom1}
\{e_{ij},e_{kl}\}=\delta_{jk} e_{i l}-\delta_{l i} e_{kj} \qquad (1\leq i,j,k,l \leq n) 	\, ,
\end{equation}
and the Cartan subalgebra $\mathcal{h}$ being determined by  
\begin{equation}\label{kom1}
[e_{i,i+1},e_{i+1,i}]=e_{i,i}-e_{i+1,i+1}:=h_i \qquad (1\leq i \leq n-1) 	\, ,
\end{equation}
which are easily verified to satisfy the Poisson-Jacobi relations. A general description of the monomials for generic $n$ can be found in \cite{cam23}. For $n=3$, the dimension of the centralizer $C_{S(\mathfrak{sl}(3))}(\mathfrak{h} )$ is seven, and an explicit basis can be chosen naturally as $\mathcal{B}=\left\{h_1, h_2, p_{1,2}, p_{1,3}, p_{2,3}, p_{1,2,3}, p_{1,3,2} \right\}$. However, only six of the elements are algebraically independent, as the following relation holds:
\begin{equation}\label{alde1}
p_{1,2}p_{1,3} p_{2,3}-p_{1,2,3} p_{1,3,2}=0.
\end{equation}
However, it is important to observe that no generator can be skipped from the basis, and the algebra needs to be understood as a Poisson algebra with an additional relation. This relation is an analogue of the closure relation observed for the Racah algebra, but in this abstract framework of enveloping algebras of Poisson-Lie algebras. This commutant can be used to define both the Hamiltonian, the integrals and the symmetry algebra in an algebraic way. The algebra $\mathcal{A}_3$ can be presented in different forms, using different symmetry properties of the indices. A suitable choice for the generators of the Poisson algebra is given by 
\begin{equation}
c_i:=\frac{1}{3}\sum_{j=1}^2 (3-j)h_j-\sum_{j=1}^{i-1} h_j \qquad c_{ij}:=p_{i,j},
\end{equation} 
\begin{equation}
f_{ijk}:=\frac{1}{2}(p_{i,k,j}-p_{i,j,k})  \qquad g_{ijk}:=\frac{1}{2}(p_{i,k,j}+p_{i,j,k}) \, 
\end{equation}
where $i \neq j \neq k \in \{1,2,3\}$. The polynomials display different symmetry properties $c_{ij}=c_{ji}$ (symmetric), $f_{ijk}=-f_{jik}=f_{jki}$ (skew-symmetric) and $g_{ijk}=g_{jik}=g_{jki}$ symmetric. The construction of the Poisson algebra relies on the relation of the underlying $\mathfrak{sl}(3)$ algebra. In particular, it can be shown that the generators lead to the following algebra with $\{c_i, \cdot\}=0$
\begin{equation}\label{equafin} 
\begin{split}
\{c_{ij},c_{jk}\}&=2 f_{ijk}\\
\{c_{jk},f_{ijk}\}&=(c_{ik}-c_{ij})c_{jk}+(c_j-c_k)g_{ijk}\\
\{c_{jk},g_{ijk}\}&=(c_j-c_k)f_{ijk}\\
\{f_{ijk},g_{ijk}\}&=\frac{1}{2}\bigl((c_i-c_k)c_{ij}c_{jk}+(c_k-c_j)c_{ki}c_{ij}+(c_j-c_i)c_{jk}c_{ki} \bigl) 
\end{split}
\end{equation}
The corresponding Casimir invariants and their properties were also studied. In addition, it was demonstrated how for a specific realization which has similarity with the Marsden-Weinstein approach, the algebra $A_n$ reduces to $R(n)$, where, in particular, $\mathcal{A}_3$ becomes $R(3)$. In this short paper, we will be concerned with the quantization of the algebra $\mathcal{A}_3$ and the explicit form of the polynomial algebra involving the commutator. There will be additional correction terms, as the symmetrization map relates higher order terms from the Poisson algebra and the commutator algebra. This is particularly true when elements are no longer linear but polynomials. This requires several lower order terms to be added.
 
\section{Quantization and quantum $\mathcal{A}_3$}
In this section we work in the context of Definition 1. The problem can be solved effectively using, for example, symbolic computation packages. Here we will use the results obtained in the Poisson-Lie frame and apply the symmetrization map, in order to find the quantum correction to the basis and the algebra. In the Lie algebra context, the commutator is then given by 
\begin{equation}\label{kom1}
[E_{ij},E_{kl}]=\delta_{jk} E_{i l}-\delta_{l i} E_{kj} \qquad (1\leq i,j,k,l \leq n) 	\, ,
\end{equation}
with the Cartan generators being $H_i=E_{i,i}-E_{i+1,i+1}$. 
We use the following notation for the generators of the polynomial algebra (the symbol $\hat{l}$ indicates the quantized generator of $l$)
\begin{equation}
\begin{split}
\hat{P}_{i_1,\dots ,i_d} &= E_{i_1,i_2}E_{i_2,i_3}\dots E_{i_{d-1},i_d}E_{i_d,i_1},\\
\hat{c}_i:=& \frac{1}{3}\sum_{j=1}^2 (3-j)H_j-\sum_{j=1}^{i-1} H_j \qquad \hat{c}_{ij}:= S( P_{i,j}),\\
\hat{f}_{ijk}:= & \frac{1}{2}( S(P_{i,k,j})-S(P_{i,j,k}))  \qquad \hat{g}_{ijk}:=\frac{1}{2}(S(P_{i,k,j})+S(P_{i,j,k})),
\end{split}
\end{equation}
where $i \neq j \neq k \in \{1,2,3\}$ and $S(.)$ indicates that the symmetrization map (\ref{syma}) (or Weyl ordering) has been applied to all the monomials composing the polynomials, i.e.,

\begin{equation}
 S(AB)= \frac{1}{2}(AB+BA) 
\end{equation}
\begin{equation} 
 S(ABC)=\frac{1}{6}(ABC+ACB+BAC+BCA+ CAB+CBA) 
\end{equation}
Here, all the polynomials are written in terms of the following ordered generators, where the order has also been taken into account  to perform the computation of the quantum correction: 
\begin{equation} 
 \{H_{3},H_{2},H_{1},E_{32},E_{31}, E_{21},E_{12},E_{13},E_{23}\} 
 \end{equation}
Specifically, this means that the polynomials that are written using the symmetrization map will need in the closure of the polynomial algebra to be expressed over this basis, and the ordering needs to be consistently kept 
\begin{equation} 
\hat{c}_{ij}= S(P_{ij})= \frac{1}{2} \{E_{ij},E_{ji}\} = E_{ij}E_{ji}-\frac{1}{2}(c_i-c_j) 
\end{equation}
and explicitly including the lower order quantum correction
\begin{equation}
\begin{split}
\hat{c}_{12} =& E_{12}E_{21} - \frac{1}{2} H_1,\quad 
 \hat{c}_{23} = E_{23}E_{32} - \frac{1}{2} H_2 ,\\
  \hat{c}_{13} =& E_{13}E_{31} - \frac{1}{2} H_1 - \frac{1}{2} H_{2},\quad
   \hat{f}_{ijk}= \frac{1}{2}( S(P_{i,k,j})-S(P_{i,j,k})).
\end{split}
\end{equation}
This implies that third degree polynomials adopt the expression 
\begin{equation}
\begin{split}
\hat{f}_{123}=& \frac{1}{2}E_{13}E_{31} - \frac{1}{2} E_{12}E_{23}E_{31} + \frac{1}{2} E_{13}E_{21}E_{23},\\ 
\hat{f}_{132}=& \frac{1}{2}E_{13}E_{31} - \frac{1}{2} E_{12}E_{23}E_{31} - \frac{1}{2} E_{13}E_{21}E_{23}, 
\end{split}
\end{equation}
but still satisfying the symmetry property of the initial $f_{123}$ in the variables of the dual space
\begin{equation}
\hat{g}_{ijk}= \frac{1}{2}( S(P_{i,k,j})+S(P_{i,j,k}))
\end{equation} 
The explicit expression in the ordered basis, with the addition of the first and second degree terms is 
\begin{equation}
\begin{split}
\hat{g}_{123}= \frac{1}{6} H_1 - \frac{1}{6} H_2 - \frac{1}{2} E_{12} E_{21} + E_{23} E_{32} + \frac{1}{2} E_{12} E_{23} E_{31} + \frac{1}{2} E_{13} E_{21} E_{32},\\
\hat{g}_{132}= \frac{1}{6} H_1 - \frac{1}{6} H_2 - \frac{1}{2} E_{12} E_{21} + E_{23} E_{32} + \frac{1}{2} E_{12} E_{23} E_{31} + \frac{1}{2} E_{13} E_{21} E_{32}.  
\end{split}
\end{equation}
Again, the symmetries of the initial $g_{ijk}$ in the variables of the dual space are fulfilled. If we proceed to symmetrize the terms obtained in the classical framework in the variables of the dual space, the following relations are obtained: 

\begin{equation}\nonumber
[\hat{c}_i,\hat{c}_j]=0, \quad  [\hat{c}_i,\hat{f}_{jkl}]=0 ,\quad  [\hat{c}_i,\hat{g}_{jkl}]=0 ,\quad  [\hat{c}_{ij},\hat{c}_{kl}]=0,\quad [\hat{c}_{ij},\hat{c}_{jk}]=2 \hat{f}_{ijk}, 
\end{equation}  

\begin{equation}
\begin{split}
[\hat{c}_{jk},\hat{f}_{ijk}]=& S( \hat{c}_{jk} ( \hat{c}_{ik}-\hat{c}_{ij}) + (\hat{c}_j- \hat{c}_k) \hat{g}_{ijk} ) )+ \frac{1}{2} \hat{c}_j^2  -\frac{1}{12} \hat{c}_k^2 - \frac{1}{6} \hat{c}_i \hat{c}_j + \frac{1}{6} \hat{c}_i \hat{c}_k \\
 =&  \frac{1}{2} \{\hat{c}_{jk},\hat{c}_{ik}\} + \frac{1}{2} 
\{\hat{c}_{jk},\hat{c}_{ij}\}  + \frac{1}{2} \hat{c}_j^2 -\frac{1}{12} \hat{c}_k^2 - \frac{1}{6} \hat{c}_i \hat{c}_j + \frac{1}{6} \hat{c}_i \hat{c}_k \\
[\hat{c}_{jk},\hat{g}_{ijk}] =& S(  (\hat{c}_j -\hat{c}_k)  \hat{f}_{ijk} ) = \frac{1}{2}  (  (\hat{c}_j - \hat{c}_k) \hat{f}_{ijk} + \hat{f}_{ijk} (\hat{c}_j - \hat{c}_k )  ),\\
[\hat{f}_{ijk},\hat{g}_{ijk}]=& S( \frac{1}{2}(\hat{c}_i -\hat{c}_k)\hat{c}_{ij} \hat{c}_{jk}  + \frac{1}{2}(\hat{c}_k -\hat{c}_j )\hat{c}_{ki} \hat{c}_{ij} + \frac{1}{2}(\hat{c}_j -\hat{c}_i)\hat{c}_{jk} \hat{c}_{ki} ) + \frac{1}{8} \hat{c}_i^2 \hat{c}_j \\
 &-  \frac{1}{8} \hat{c}_i^2 \hat{c}_k - \frac{1}{8} \hat{c}_j^2 \hat{c}_i + \frac{1}{8} \hat{c}_j^2 \hat{c}_k  + \frac{1}{8} \hat{c}_k^2 \hat{c}_i - \frac{1}{8} \hat{c}_k^2 \hat{c}_j 
\end{split}
\end{equation}
A long but routine computation shows that the constraints can also be determined in the quantum setting, considering the additional correction terms
\begin{equation}
\begin{split}
S&( \hat{g}_{ijk} \hat{g}_{kji}) + S(\hat{f}_{ijk} \hat{f}_{kji} ) - S( \hat{c}_{ij} \hat{c}_{jk} \hat{c}_{ki} )+ \frac{1}{6} \hat{c}_{i}^2  + \frac{1}{6} \hat{c}_i \hat{c}_j - \frac{1}{6} \hat{c}_j^2 - \frac{1}{4} \hat{c}_{ij}^2 + \frac{1}{6} \hat{c}_{ij}\hat{c}_{ik} \\
- &\frac{1}{4} \hat{c}c_{ik}^2 + \frac{1}{6} \hat{c}_{ik} \hat{c}_{jk} - \frac{1}{4} \hat{c}_{jk}^2+ \frac{1}{2} \hat{c}_{ij} \hat{c}_i \hat{c}_j + \frac{5}{4} \hat{c}_{ij} \hat{c}_i \hat{c}_j  + \frac{1}{2} \hat{c}_{ij} \hat{c}_i \hat{c}_j - \frac{1}{4} \hat{c}_{ik} \hat{c}_i^2  - \frac{1}{4} \hat{c}_{ik} \hat{c}_i \hat{c}_j \\
+& \frac{1}{2} \hat{c}_{ik} \hat{c}_j^2 + \frac{1}{2} \hat{c}_{jk} \hat{c}_i^2 - \frac{1}{4} \hat{c}_{jk} \hat{c}_i \hat{c}_j - \frac{1}{4} \hat{c}_{jk} \hat{c}_j^2 + \frac{1}{6} \hat{f}_{ijk} =0.
\end{split}
\end{equation}

\section{Conclusion} 
The purely algebraic approach developed in recent years in the context of universal enveloping algebras allows to introduce a new perspective in the construction and analysis of superintegrable systems, avoiding the explicit manipulation of differential operator algebras, replacing them by intrinsic algebraic structures contained in the enveloping algebra of Lie algebras. We have briefly reviewed the procedure, as well as how to implement it when starting from the analytical ansatz. It is hoped that this approach, besides specific applications to superintegrable systems, will also lead to new avenues for the study of Lie algebras, their enveloping algebra, as well as their representation theory. The method has been illustrated by means of the Lie algebra $\mathcal{A}_3$, where both the classical and quantum cases have been revisited, showing the deep relation between the polynomial algebras obtained from commutants in enveloping algebras and the Racah algebra $R(3)$, once an appropriate differential realisation has been chosen. Formally, the method can also be applied to higher-rank algebras, at both the classical and quantum cases, possibly providing new insights into the structure and hierarchical classification of higher-dimensional superintegrable systems. Work in this direction is currently in progress.

 \medskip 
{\bf Acknowledgements:}
This work was partially supported by the Future Fellowship FT180100099 and the Discovery Project DP190101529 from the Australian Research Council. RCS  acknowledges financial support by the research
grant PID2019-106802GB-I00/AEI/10.13039/501100011033 (AEI/ FEDER, UE).

\end{document}